\documentclass[preprint]{aastex62}
\usepackage{epsf}
\usepackage{natbib}
\usepackage{graphicx}
\usepackage{float}
\usepackage{afterpage}
\usepackage{amssymb}
\usepackage{enumitem}
\usepackage{bm,bbm,amssymb, amsmath}
\usepackage{enumerate}
\usepackage{url}
\usepackage[section]{placeins}

\usepackage{mathrsfs}
\usepackage{ulem}

\usepackage{hyperref}

\begin{document}
\title{\Large Dust grains cannot grow to millimeter sizes in protostellar envelopes}
\correspondingauthor{Kedron Silsbee}
\email{kpsilsbee@utep.edu}

\author[0000-0003-1572-0505]{Kedron Silsbee}
\affil{Max-Planck-Institut f\"ur
Extraterrestrische Physik, 85748 Garching, Germany}
\affil{University of Texas at El Paso, El Paso, TX, 79968, USA}

\author[0000-0002-4324-3809]{Vitaly Akimkin}
\affil{Institute of Astronomy, Russian Academy of Sciences, 48 Pyatnitskaya St., Moscow 119017, Russia}

\author[0000-0002-4324-3809]{Alexei V. Ivlev}
\affil{Max-Planck-Institut f\"ur
Extraterrestrische Physik, 85748 Garching, Germany}

\author[0000-0003-1859-3070]{Leonardo Testi}
\affil{European Southern Observatory, Karl-Schwarzschild-Str. 2, D-85748 Garching bei M\"unchen, Germany}
\affil{INAF-Osservatorio Astrofisico di Arcetri, Largo E. Fermi 5, I-50125 Firenze, Italy}

\author[0000-0003-1613-6263]{Munan Gong}
\affil{Max-Planck-Institut f\"ur
Extraterrestrische Physik, 85748 Garching, Germany}

\author[0000-0003-1481-7911]{Paola Caselli}
\affil{Max-Planck-Institut f\"ur
Extraterrestrische Physik, 85748 Garching, Germany}

\begin{abstract}
A big question in the field of star and planet formation is the time at which substantial dust grain growth occurs.  
The observed properties of dust emission across different wavelength ranges have been used as an indication that millimeter-sized grains are already present in the envelopes of young protostars.
However, this interpretation is in tension with results from coagulation simulations, which are not able to produce such large grains in these conditions.  In this work, we show analytically that the production of millimeter-sized grains in protostellar envelopes is impossible under the standard assumptions about the coagulation process.  We discuss several possibilities that may serve to explain the observed dust emission in the absence of in-situ grain growth to millimeter sizes.
\end{abstract}
\section{Introduction}
One of the key steps in the formation of planets is the growth of dust particles from the sub-micron sized grains in the interstellar medium to pebble-sized aggregates. It is generally thought that this growth occurs in protoplanetary disks, which indeed show widespread evidence for the presence of large grains \citep[e.g.][]{Testi14}. Nevertheless, evidence has accumulated in the last decade, suggesting that dust evolution begins already in clouds and cores (before the formation of protoplanetary disks), where infrared observations have convincingly shown the presence of a significant population of grains grown to a few microns \citep{Pagani10}. More recently, millimeter observations of the spectral index of dust emission appear to be consistent with the presence of millimeter-sized grains in the inner region of class 0 protostellar envelopes
\citep[e.g.][]{Kwon09,Miotello14,Galametz19}. These latter results are surprising, because taken at face value, they appear to be at odds with the expected timescales for grain growth in cores and envelopes of protostars.  Simulations of grain coagulation typically do not predict growth beyond tens of microns \citep[e.g.][]{Ossenkopf93,Ormel09}.  \citet{Galametz19} and \citet{Agurto-Gangas19} have shown that the derivation of the grain growth constraints from the spectral index is not always straightforward, and that not all protostars show definitive evidence for grain growth. Nonetheless, through a careful analysis of a large sample of protostellar envelopes, \citet{Galametz19} show evidence of the possible presence of millimeter-size grains in the inner $\sim$500~au of at least some protostellar envelopes. Indirect evidence from mm polarization observations seem to confirm the inferred presence of grains of at least 10 $\mu$m \citep[see e.g.][]{Valdivia19}.  It is thus important to revisit the modeling of grain growth in cores and protostars, with the aim of understanding whether growth to millimeter sizes in these environments is indeed possible under certain circumstances.
\par

    There have been many studies of dust coagulation in molecular clouds and prestellar cores.  \citet{Rossi91} considered the growth of dust grains via coagulation and condensation of gas-phase molecules in a cloud undergoing pressure-free spherical collapse.  They assumed the grain velocities to arise in response to gas turbulence.  Because of the rapid collapse in their model and the fact that the possibility of fractal growth/high grain porosity was not considered, they were only able to grow grains of a few microns in size.
    A wider range of possible sources of collision velocity (turbulence, Brownian motion, and motions arising from grain asymmetries and differential gravitational settling) was considered by \citet{Ossenkopf93}.
    He found the dominant driver of coagulation to be turbulence at number densities below $10^8$~cm$^{-3}$, and Brownian motion at higher densities.  He did not run the simulations for long enough to see the growth of large grains.
    \citet{Ormel09} considered collision velocities arising in response to turbulence in the gas, and developed a detailed model for collision outcomes including sticking, fragmentation and compaction.  They considered evolution timescales as long as $10^7$ years at densities of $10^7$~cm$^{-3}$ and were able to grow grains up to 0.6 mm in size, with the maximum size determined by fragmentation.
    Recently, \citet{Silsbee20} and \citet{Guillet20} considered an additional mechanism leading to relative velocity --- the differential coupling of grains of different sizes to the neutral and magnetized component of the gas.  They both find that it results in a very rapid removal of the smallest grains from the size distribution, but has little effect on the maximum grain size.
\par
However, there are several uncertain parameters in grain growth models related to the dust physics.  Their variation has not been systematically explored in the dust coagulation simulations, partly due to the difficulty of covering a large parameter space. For example,
the evolution of the dust porosity is a matter of substantial debate in the literature \citep[e.g.][]{Shen08, Wada08, Ormel09, Okuzumi09}, with uncertainty as to whether the grains remain compact, or undergo fractal growth.  There is also a significant uncertainty in the critical collision velocity required for grain fragmentation, with different theoretical and experimental work pointing to different threshold velocities from centimeters per second to several tens of meters per second, depending on grain composition and size \citep{Blum08, Ormel09, Wada13, Kimura15, Gundlach15, Musiolik16}.
Moreover, while much work has assumed a Kolmogorov slope of the turbulent energy spectrum, this need not necessarily be the case.  In protoplanetary disks, turbulence driven by the magnetorotational instability was shown to have a shallower power spectrum \citep{Gong20}.  Furthermore, the rate of dust growth was shown to be very sensitive to the slope of the turbulent power spectrum \citep{Gong21}.  
\par
In this paper, we consider a simple, but general coagulation model that can be solved analytically.  By using the analytic solution, we are able to put rigorous upper bounds on the growth that may take place, without having to numerically explore the many-dimensional space of parameters that affect the coagulation rate.  We consider the degree of compaction, the maximum collision velocity that can be sustained without fragmentation, and the normalization and slope of the turbulent power spectrum as free parameters.

\section{Calculation of the minimum grain growth timescale}
While there is some non-trivial dependence of the optical properties on grain porosity, it is reasonable to approximate that the optical properties of a grain are a function of the product of its size and filling factor \citep[][see their Figure 11]{Kataoka14}, that is
\begin{equation}
a_{\rm opt} = \phi a,
\label{eq:aopt}
\end{equation}
where $\phi<1$ is the fraction of the grain volume occupied by solid material and $a$ is the grain radius.  For grain growth to greatly affect the spectral index at wavelength $\lambda$, the grains should have $|m-1|a_{\rm opt} \gtrsim \lambda/(2\pi)$, where $m$ is the complex refractive index.
\par
Let us define $\tau_{\rm coll}$ as the expected time for a grain to double its mass.  We note from the simulations performed in \citet[][see top panels of their Figure 7]{Silsbee20}, that, at least for the Kolmogorov turbulence spectrum and compact grains, the grain size distribution remains relatively narrow, even as growth proceeds.  We therefore approximate that at any given time, the size distribution is monodisperse [cf. Appendix A in \citet{Ormel09}].  We verify in Appendix \ref{appendA} that this approximation introduces very little error when compared with simulations of the full size distribution.  Therefore, $\tau_{\rm coll}$ is exactly equal to the expected time for an individual grain to experience a collision:
\begin{equation}
\tau_{\rm coll} = \frac{1}{n_d \sigma v_{\rm coll}},
\label{eq:tau0Simp}
\end{equation}
where $\sigma$ is the collision cross-section, $v_{\rm coll}$ the collision velocity, and the number density $n_d$ can be easily related to 
the molecular hydrogen number density $n_g$, 
the grain size $a$, dust-to-gas ratio $f_d$, and size-dependent dust density $\rho_d(a)$:
\begin{equation}
n_d = \frac{3 f_d \rho_g}{4 \pi a^3 \rho_d(a)},
\end{equation}
where $\rho_g = n_g \mu_g$ and $\mu_g$ is the mean mass per hydrogen molecule.  We assumed a gas with 5 parts molecular hydrogen to one part helium, so $\mu_g = 2.8 m_{\rm H}$, with $m_{\rm H}$ the mass of a hydrogen atom.  Throughout this work, we assume spherical grains.  Also, we assume a geometric cross section $\sigma = 4 \pi a^2$, as Coulomb focusing has been shown to be unimportant except for small grains with very small encounter velocities \citep{Akimkin20}.
\par
As the grains grow, they may not necessarily remain compact.  We parameterize this by a fractal dimension $D > 2$, such that the grain density satisfies
\begin{equation}
\rho_d(a) = \rho_s \left(\frac{a}{a_0}\right)^{D-3},
\label{eq:rho_d}
\end{equation}
where $D=3$ corresponds to compact grains with constant solid density $\rho_s$, and $D\rightarrow2$ to fractal grains with
constant mass to surface area ratio.  $a_0$ is the initial grain size.
Noting that the filling factor $\phi = \rho_d/\rho_s$, Equations \eqref{eq:aopt} and \eqref{eq:rho_d} imply the following relation between size $a$ and the optical equivalent size:
\begin{equation}
    \left(\frac{a}{a_0}\right)^{D-2} = \frac{a_{\rm opt}}{a_0}.
    \label{eq:a-aopt}
\end{equation}
\par
We consider collision velocities arising in response to turbulence in the gas, as these are thought to dominate for grains larger than $0.1~\mu$m \citep{Silsbee20}.  This turbulence need not necessarily have a Kolmogorov power spectrum.  We let $p$ be the slope of the power spectrum of the turbulence (5/3 for Kolmogorov turbulence).  \citet{Gong21} showed [see Equation~(30) therein; note that for equal particle sizes, the prefactor of the non-vanishing term in that equation should include an additional factor of 2] that, if both grains have Stokes number $ {\rm St} \ll 1$, and the eddy correlation time is equal to the eddy turnover time, we can write the collision velocity between two grains of equal sizes as 
\begin{equation}
v_{\rm coll} = v_g \psi(p) {\rm St}^q,
\label{eq:generalVelocity}
\end{equation}
where $v_g$ is the turbulent velocity, $\psi(p) = \sqrt{8(p-1)/(p+1)}$ is a factor of order unity, and $q = (p-1)/(3-p)$; for all turbulence models with $1 < p < 2$, we have $0 < q < 1$.  The Stokes number ${\rm St} = \tau_s/\tau_L$, where $\tau_L = r/(v_g \sqrt{p-1})$ is the eddy turnover time at the injection scale $r$, and $\tau_s(a) = a\rho_d(a)/(\rho_g v_{\rm th})$ is the grain stopping time due to gas friction in the Epstein regime \citep{Epstein1924}.  Here, $v_{\rm th} = 0.92 \sqrt{4k_BT/{\pi m_{\rm H}}}$ is the mean thermal velocity of the H$_2$ particles, with a correction factor of 0.92 to account for the presence of helium.
\par
Combining Equations \eqref{eq:tau0Simp} through \eqref{eq:generalVelocity}, we find
\begin{equation}
\tau_{\rm coll}(a) = \frac{a \rho_d(a)}{3 f_d n_g \mu_g v_g \psi(p)} \left(\frac{n_g \mu_g v_{\rm th} r}{a \rho_d(a) v_g \sqrt{p-1}}\right)^q.
\label{eq:tau0}
\end{equation}
To solve for the time evolution of the size distribution, we note that $\dot a/a = \frac{1}{D} \dot m/m$.  We can then write 
\begin{equation}
\frac{da}{dt} = \frac{a}{D \tau_{\rm coll}(a)}.
\end{equation}
Substituting Equation \eqref{eq:tau0} for $\tau_{\rm coll}$, we obtain the solution 
\begin{equation}
a(t) = a_0 \left[\frac{(1-q)(D-2)t}{D \tau_{\rm coll}(a_0)} + 1 \right]^\frac{1}{(1-q)(D-2)}.
\label{eq:growthEquation}
\end{equation}
Using Equation \eqref{eq:growthEquation}, we can solve for the growth time $\tau_{\rm gr}$ at which the grain reaches size $a$: 
\begin{equation}
\tau_{ \rm gr} = \frac{D \tau_{\rm coll}(a_0)}{(1-q)(D-2)} \left[\left(\frac{a}{a_0}\right)^{(1-q)(D-2)} -1\right] \leq \tau_{\rm env},
\label{eq:tau}
\end{equation}
where $\tau_{\rm env}$ is the lifetime of the envelope. Note that for $D \rightarrow 2$ the grain growth becomes exponential, $a(t) = a_0 e^{t/2\tau_{\rm coll}(a_0)}$, and then the condition in Equation~\eqref{eq:tau} is reduced to $2\tau_{\rm coll}(a_0)\ln(a/a_0)\leq \tau_{\rm env}$.
\par
If the collision velocities are too high, then the grains may shatter instead of sticking when they collide.  We parameterize this by assuming that there is a (size-independent) maximum collision velocity $v_{\rm max}$ above which grain fragmentation occurs.  The highest speed collisions will be between the largest grains.  Because of this, and the fact that the threshold velocity for sticking is generally thought to decrease with size \citep[see e.g.][]{Wada13}, $v_{\rm max}$ may be thought of as the threshold velocity for the largest grains.  Requiring the collision velocity between two grains of size $a$ to be less than $v_{\rm max}$ gives the relation
\begin{equation}
v_g \psi(p) \left(\frac{a_{\rm opt} \rho_s  v_g \sqrt{p-1}}{n_g \mu_g v_{\rm th} r} \right)^q \leq v_{\rm max}.
\label{eq:v}
\end{equation}

\subsection{Fiducial parameters}
\label{sect:fiducialParams}
In what follows, we keep our derivations fully general, but we do provide a few numerical estimates and a figure.  These are made for the specific parameters listed in Table \ref{tab:fiducialParams}.
\begin{table}[h!]
  \begin{center}
    \caption{{\bf Fiducial Parameters}}
    \label{tab:fiducialParams}
  \begin{tabular}{l l l }
        symbol & meaning & fiducial value \\
        \hline
     $r$ & length scale of environment & 1000~AU \\
     $\tau_{\rm env}$ & lifetime of Class 0 stage & $10^5$~yr\\
     $p$ & slope of turbulent energy spectrum & 5/3 \\
     $q$ & $(p-1)/(3-p)$ & 1/2 \\
     $T$ & temperature & 20~K\\
     $\mu_g$ & mean mass per hydrogen molecule & $2.8 m_{\rm H}$ \\
     $n_g$ & number density of hydrogen molecules & $10^7$~cm$^{-3}$\\
     $\rho_s$ & dust material density & 3~g~cm$^{-3}$\\
     $f_d$ & dust-to-gas mass ratio & 0.01 \\
     $a_0$ & initial grain size & 1~$\mu$m \\
     $D$ & fractal dimension of grain growth & 3 \\
     $v_{\rm max}$ & grain fragmentation velocity & 10~m/s 
    \end{tabular}
  \end{center}
\vspace{-.5cm}
\end{table}
\subsection{Results}
Based on the observations of low spectral index of the dust emission between 1.3 and 3.2 mm shown in \citet{Galametz19}, we are interested in growing grains that have $a_{\rm opt}$ around 1 mm. Equations \eqref{eq:tau} and \eqref{eq:v} define two regions in the plane of $v_g$ and $n_g$ which are ``acceptable" in the sense of having sufficiently rapid coagulation without introducing excessive collisional velocities.  Equations \eqref{eq:tau} and \eqref{eq:v} can be written in the form
\begin{equation}
\mathscr{G} v_g^{-1-q} n_g^{q-1} \leq 1, \quad \quad \mathscr{F} v_g^{1+q} n_g^{-q} \leq 1,
\label{eq:simplifiedInequalities}
\end{equation}
where 
\begin{align}
& \mathscr{G} = \frac{a_0 \rho_s D}{3 \mu_gf_d \tau_{\rm env} \psi(p)(1-q)(D-2)} \left(\frac{\mu_g v_{\rm th}r}{a_0 \rho_s \sqrt{p-1}}\right)^q 
\nonumber\\
&
\times \left[\left(\frac{a_{\rm opt}}{a_0}\right)^{1-q}-1\right],
\end{align}
and 
\begin{equation}
\mathscr{F} = \frac{\psi(p)}{v_{\rm max}} \left(\frac{a_0\rho_s \sqrt{p-1}}{\mu_gv_{\rm th}r}\right)^q \left(\frac{a_{\rm opt}}{a_0}\right)^{q}.
\end{equation}
 \begin{figure}
\centering
\includegraphics[width=0.9\columnwidth]{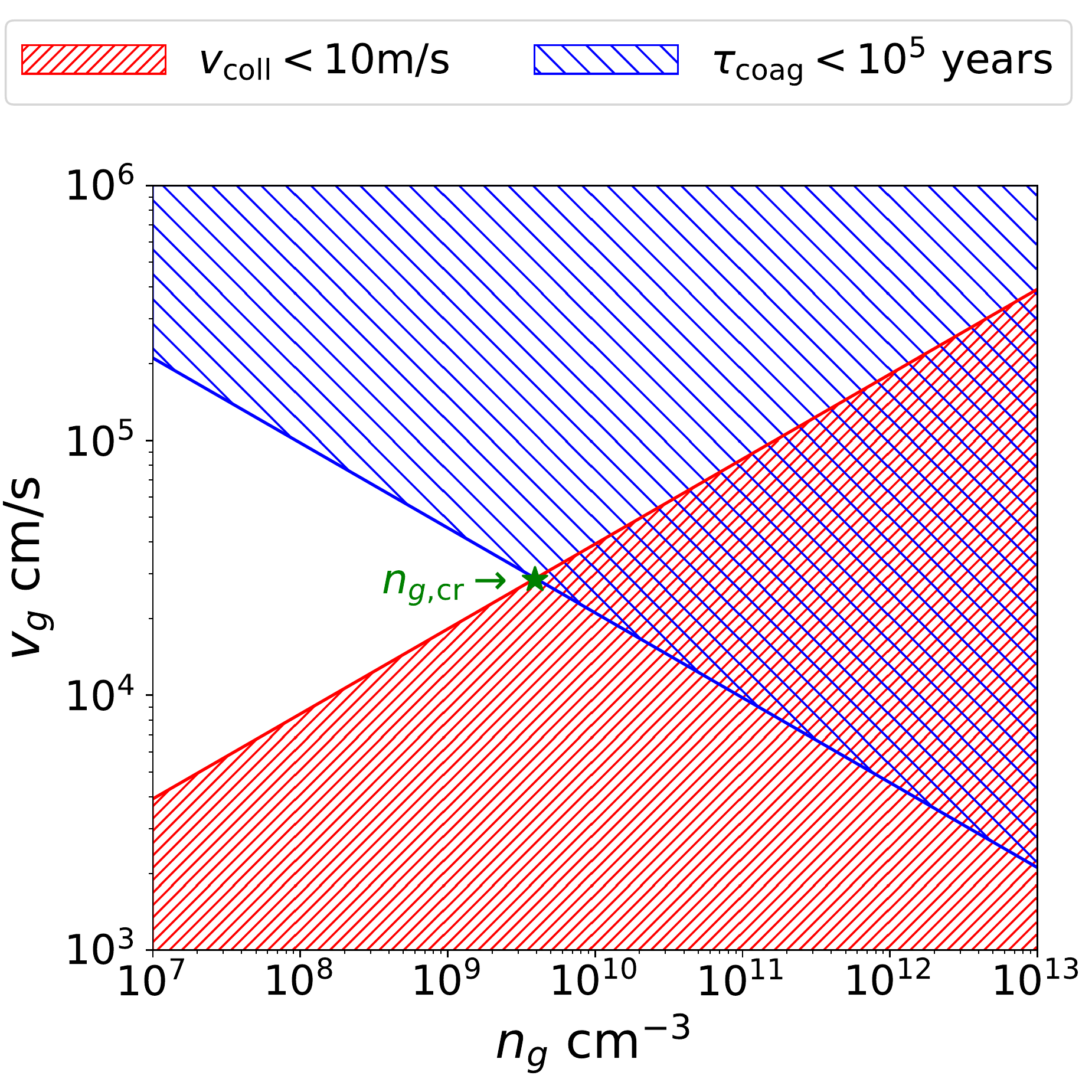}
\caption{Regions of $n_g - v_g$ space in which the maximum collision velocity is below $v_{\rm max}$ [red shading; see Equation \eqref{eq:v}], and the timescale for grain growth to 1 mm is below $\tau_{\rm env}$ [blue shading; see Equation \eqref{eq:tau}]. This is made assuming the fiducial parameters in Table \ref{tab:fiducialParams}.  The region with overlapping red and blue shading is the allowed parameter space for grain growth to 1 mm.  The star corresponds to the critical density above which this is possible (see Equation \eqref{eq:ng}).}
\label{fancyInequalities}
\end{figure}

  Equations \eqref{eq:simplifiedInequalities} are plotted in Figure \ref{fancyInequalities}.  The requirement that $v_{\rm coll} \leq v_{\rm max}$ provides an upper bound on $v_g$, and the requirement that $\tau_{\rm gr} \leq \tau_{\rm env}$ provides a lower bound on $v_g$.  At low density, we cannot simultaneously satisfy these requirements, but above some density the regions overlap.  For our fiducial parameters, Figure \ref{fancyInequalities} shows that this critical density is about $4 \times10^9$ cm$^{-3}$.  We note that the corresponding value of the turbulent velocity $v_g$ is a very reasonable $3 \times 10^4$~cm/s, which corresponds to an isothermal Mach number of 1.1 
  \par
  The doubly-shaded region which satisfies both Equations \eqref{eq:tau} and \eqref{eq:v} exists to the right of the intersection point of the two curves.  Replacing the ``$\leq$" signs in Equations \eqref{eq:simplifiedInequalities} with ``=" and multiplying the two equations together yields the critical value $n_{g,{\rm cr}} = \mathscr{FG}$ of gas density at the intersection point:
  \begin{align}
  &n_{g,{\rm cr}} = \frac{a_0\rho_s D}{3 \mu_g f_d \tau_{\rm env} (1-q)(D-2) v_{\rm max}}\left[\frac{a_{\rm opt}}{a_0} - \left(\frac{a_{\rm opt}}{a_0} \right)^{q}\right]
  \nonumber\\
  &= 3.9 \times 10^9 \left(\frac{a_{\rm opt}}{1~{\rm mm}}\right) \left(\frac{10^5~{\rm yr}}{\tau_{\rm env}}\right) \left(\frac{10~{\rm m/s}}{v_{\rm max}}\right)~{\rm cm}^{-3},
  \label{eq:ng}
  \end{align}
where in the numerical estimate we ignore the small second term in the square brackets and take the fiducial parameters from Table~\ref{tab:fiducialParams}.  Remarkably, the slope $p$ of the turbulent power spectrum has little effect on the critical value of $n_g$.  This is because of the condition that the collision speed of the largest bodies be no more than $v_{\rm max}$.  For all the values of $p$ that we consider, the growth timescale $\tau_{\rm coll}(a)$ increases with $a$ [see Equation \eqref{eq:tau0}].  For this reason, the total growth timescale is dominated by the time taken by the final few doublings in size.  The collision velocity for these sizes will be of the order of $v_{\rm max}$, no matter what value of $p$ we take (since we are bound by the condition that $v_{\rm coll} = v_{\rm max}$ at the maximum size).  A shallower spectrum means that this maximum size is achievable with a lower value of $v_g$, but since we assume no prior knowledge of $v_g$ this does not affect our result.  For our fiducial parameters, and $a_{\rm opt}$ deduced from the numerical estimate in Equation \eqref{eq:ng}, with $n_{g,{\rm cr}} = n_g$ the implied value of $v_g$ is 210~m/s.  For steeper spectra of the turbulence (larger values of $p$), the required values of $v_g$ may become unrealistically large.  This simply means that in this case, the grain growth will proceed even more slowly than our calculations would suggest.
 \par
Furthermore, the critical density is insensitive to changes in the fractal dimension $D$, unless $D$ becomes close to 2, in which case $n_{g, {\rm cr}}$ diverges.  One can also think of a situation where $D = 3$, but the filling factor $\phi$ is much less than unity. In this case $\rho_d/\rho_s= a_{\rm opt}/a= \phi$, and we readily derive the expression for $n_{g,{\rm cr}}$ which coincides with Equation~\eqref{eq:ng} for $D=3$ (again, ignoring the second term in the square brackets).  
 
\subsection{Maximum grain size}
\label{sect:maxGrainSize}
 Assuming $a_{\rm opt} \gg a_0$ we can drop the second term in parentheses in Equation \eqref{eq:ng}, and solve for $a_{\rm opt}$, leaving us with 
\begin{align}
&a_{\rm opt} = \frac{3 \mu_g f_d (1-q)(D-2)}{\rho_s D} n_g \tau_{\rm env} v_{\rm max} 
\nonumber \\ 
&= 2.5\left(\frac{n_g}{10^7~{\rm cm}^{-3}}\right)\left(\frac{\tau_{\rm env}}{10^5~{\rm yr}}\right) \left(\frac{v_{\rm max}}{10~{\rm m/s}}\right)~\mu{\rm m}.
\label{eq:finalResult}
\end{align}
The numerical estimate assumes the fiducial value of all well-constrained parameters that are not explicitly listed.  Since $v_g$ could be smaller than the maximum value permitted by the condition $v_{\rm coll} \leq v_{\rm max}$, Equation \eqref{eq:finalResult} is an upper bound on the possible grain growth.  
\par
Based on a study of relative abundances of different stages of protostars, \citet{Evans09} estimated a mean lifetime for class 0 protostars of $\tau_{\rm env} = 1.0 \times 10^5$ years. We note that this provides an upper limit on the time available for coagulation in this phase.  The existence of an envelope for $10^5$ years does not necessarily imply that the {\it same} material is present in the envelope for the whole time.  Due to infall motion, it may be that the material currently present in the observed envelopes has been at that density for much less than $10^5$ years, and would thus display even less dust growth. On the other hand, one could imagine that some growth has taken place in the prestellar phase, but this effect is unlikely to be significant (see section \ref{sect:prestellarGrowth}).
  \par
  \citet{Kristensen12} modelled the dust emission in a sample of fifteen class 0 sources.  Based on their modelling, at a distance of 500 AU from the central star, the mean density is $n_g = 7 \times 10^6$~cm$^{-3}$ with a range from $1 \times 10^6$~cm$^{-3}$ to $2 \times 10^7$~cm$^{-3}$.  Following \citet{Birnstiel12}, we take $v_{\rm max} = 10$~m/s.
  \par
  Using the fiducial parameters given in Section \ref{sect:fiducialParams} we find $a_{\rm opt} = 2.5~\mu$m.  One could imagine a factor of a few change in $f_d/\rho_s$ coming from the effect of freeze-out (see Section \ref{subsect:freezeout}), but even taking this into account, we would have to increase the product $n_g \tau_{\rm env} v_{\rm max}$ by two orders of magnitude to account for mm-sized grains in these cores.
 \par
Equation \eqref{eq:generalVelocity} is derived assuming that the Stokes numbers remain small.  For $10^5$~cm$^{-3} \leq n_g \leq 10^9$~cm$^{-3}$, $10^4$~years $\leq \tau_{\rm env} \leq 10^6$~years, $2.1 \leq D \leq 3.0$ and $1.05 \leq p \leq 1.95$ (and other parameters taken from Table \ref{tab:fiducialParams}) we calculated the maximum grain size using Equation \eqref{eq:finalResult}, and then calculated the Stokes number for grains of that size.  For all cases in which the first term in brackets in Equation \eqref{eq:ng} is more than twice the second (so that dropping the second in favor of the first, as we do in Equation \eqref{eq:finalResult}, is reasonable), we find that the Stokes number for grains of size $a_{\rm opt}$ is less than $5 \times 10^{-3}$, thus justifying this approximation.
\par
It is somewhat counter-intuitive that the maximum grain size [Equation \eqref{eq:finalResult}] is almost independent of $q$ (or equivalently $p$, the slope of the turbulent energy spectrum).  In contrast, it was found in \citet{Gong21} that the coagulation rate is highly sensitive to this parameter.  This apparent discrepancy arises due to different assumptions about what is held fixed as $p$ varies.  In \citet{Gong21}, the value of $\alpha$, which determines the turbulent velocity is a fixed quantity.  In their work, the turbulent velocity is assumed to be low enough that fragmentation can be ignored.  In contrast, in the present work, since we are trying to derive an upper bound on the dust size, we are implicitly varying $v_g$ as we vary $p$,  so that Equation \eqref{eq:v} remains satisfied.

\section{Discussion}
In this section we explore some of the uncertainties in our model, and discuss the plausibility of some alternate hypotheses not involving large dust grains, to explain the observed low spectral indices. 

\subsection{Model uncertainties}

Here we analyze the possible impact of some poorly constrained processes neglected in the above consideration.

\subsubsection{Possible effects of freeze-out}
\label{subsect:freezeout}
    
    The observations motivating this paper correspond to distances from the central star of a few hundred to few thousand AU \citep{Kwon09, Miotello14, Galametz19}.  The temperature at these distances varies significantly, depending on the assumed stellar luminosity and dust distribution.  \citet{Galametz19} find temperatures ranging from 28~K to 62~K at 200~AU and from 11~K to 25~K at 2000~AU.  At these temperatures, we expect that at least some of the volatile material will be frozen out on the surface of the dust grains.  This means that the typically used dust density of 3 g cm$^{-3}$ and dust-to-gas ratio 0.01 may no longer be appropriate to use.  The total mass of dust can be obtained by adding up the contribution from all metals excepting the noble gases.  Using a solar metallicity, this gives a maximum dust-to-gas ratio of 1.2\% \citep{Asplund09}.  This practically coincides with our fiducial value of 1.0\%.  The additional material likely has a density on the order of 1~g~cm$^{-3}$.  It therefore seems that the presence of icy mantles will only affect the combination $f_d/\rho_s$ that enters in the maximum value of $a_{\rm opt}$ in Equation \eqref{eq:finalResult} by a factor of $\sim 3$ at most.
    \par
    The degree of freeze-out may also have a substantial effect on $v_{\rm max}$.  There is a wide variety of estimates in the literature for the critical fragmentation velocity, depending on grain composition and size.  It is generally thought based on both experiment \citep[e.g.][]{Gundlach15} and theoretical studies \citep[e.g.][]{Wada08} that the presence of icy mantles increases the stickiness of grains \citep[but see][]{Kimura20}.  However, we have already used a somewhat generous value of $v_{\rm max} = 10$~m/s, as adopted in \citet{Birnstiel12}, justified there as appropriate for ice-coated grains.
    \par
    In addition to any effect on the coagulation process, icy mantles covering refractory grains in dense and cold regions may dramatically change the dust opacity. The commonly used opacity model of \citet{Ossenkopf94} predicts a broad absorption feature in the far-IR range, associated with lattice vibrations of the water ice and leading to significant opacity variations at wavelengths of up to $\approx200~\mu$m. The form of the opacity function generally depends on the actual ice composition. While there are very few direct measurements of the optical constants of astrophysical ice analogs in the (sub)mm range \citep[see][]{Allodi14, Giuliano19}, these studies indicate the presence of localized, strong absorption features at such wavelengths and, hence, suggest possible opacity bumps. Additional measurements of the optical constants in this spectral range are needed to quantify the importance of absorption features and their potential contribution into the opacity index. 

\subsubsection{Coagulation prior to the Class 0 phase}
\label{sect:prestellarGrowth}
While the present work considers coagulation that occurs during the Class 0 phase, it is possible that some growth begins even prior to this stage.  Dust growth in a collapsing cloud was discussed first in \citet{Hirashita09}.  They used a free-fall model for the collapse, and studied the growth assuming the grain velocities to be determined from Brownian motion.  Because Brownian motion is only efficient for the smallest grains, they found that grains remained smaller than a micron, even at densities as high as $10^{13}$ cm$^{-3}$.  
\par
\citet{Guillet20} treated the collapse in a similar way as \citet{Hirashita09}, but included also the effects of ambipolar diffusion and turbulence on the collision velocities.  The addition of these other sources of motion (in particular turbulence) allowed the formation of $\sim 10~\mu$m grains at the time when the density reached $n_g = 10^{12}$~cm$^{-3}$.  But based on this result, it seems 
that formation of large grains during the prestellar phase is impossible except in the very dense material (which presumably becomes part of the star, rather than the envelope).  \citet{Silsbee20} analysed dust growth in the prestellar stage at a density of $n_g = 7 \times 10^{5}$~cm$^{-3}$, assuming a cloud lifetime set by ambipolar diffusion (calculated to be 10 times the free-fall time at that density).  Even in their most optimistic model, the grain sizes were capped at 10 microns.
\par
\citet{Bate22} considered the same problem of grain growth in a collapsing prestellar core.  He considered relative velocities arising from Brownian motion, pressure gradients and turbulence.  In contrast to \citet{Hirashita09} and \citet{Guillet20}, the grain coagulation was coupled to a 3D SPH simulation of the gas dynamics.  Even with this added degree of realism, grain growth at moderate densities was minimal, with the grain size not reaching 10 $\mu$m below a number density of $10^{14} - 10^{15}$ cm$^{-3}$.    The lifetime of prestellar cores at a particular density has been shown observationally to be a few times the free-fall time at that density, i.e., proportional to $n_g^{-1/2}$ \citep{Konyves15}.  Since, all else equal, the maximum coagulation rate is proportional to $n_g$, the amount of grain growth per logarithmic interval in density is dominated by the highest densities.  From these considerations, and the fact that the formation of mm-sized grains is relatively insensitive to the initial size, we consider it unlikely that grain growth in the prestellar phase would significantly alter our conclusions, unless there are mechanisms --- such as protostellar outflows \citep{Wong16, Tsukamoto21} --- that eject grains in large amounts from the very dense central region. 

\subsection{Alternative explanations for the observations}
In this section, we discuss the plausibility of several different mechanisms that could produce the observed spectral indices without requiring the presence of mm-sized grains

\subsubsection{Effects of temperature and optical depth}
The presence of large dust grains is inferred from measurements of the spectral index $\alpha$ of dust emission.  For two frequencies $\nu_1$ and $\nu_2$ with corresponding fluxes $F_{\nu_1}$ and $F_{\nu_2}$, $\alpha$ is given by  
    \begin{equation}
         \alpha = \frac{\ln{F_{\nu_1}} - \ln{F_{\nu_2}}}{\ln{\nu_1} - \ln{\nu_2}}.
         \label{eq:alpha}
    \end{equation}
   In general, the intensity emitted at frequency $\nu$ by a column of material at temperature $T$ is given by 
   \begin{equation}
       I_\nu = B_\nu(T) (1-e^{-\tau_\nu}),
   \end{equation}
   where $B_\nu(T)$ is the Planck function, and $\tau_\nu$ the optical depth at frequency $\nu$.  The spectral index $\alpha$ is then given by 
   \begin{equation}
       \alpha \equiv \frac{d \ln I_\nu}{d \ln \nu} = \frac{d \ln B_\nu}{d \ln \nu} + \frac{\tau_\nu e^{-\tau_\nu}}{1 - e^{-\tau_\nu}}\left(\frac{d \ln{\tau_\nu}}{d \ln{\nu}}\right) = \alpha_{\rm Pl} + \frac{\beta\tau_\nu }{e^{\tau_\nu}-1},
       \label{eq:fullAlpha}
   \end{equation}
 where $\alpha_{\rm Pl}$ is the spectral index of the Planck function and $\beta$ is the logarithmic slope of the dust opacity.     For optically thin emission Equation \eqref{eq:fullAlpha} reduces to the relation $\alpha = \alpha_{\rm Pl} + \beta$. 
 \par
 The observations of the spectral index presented in \citet{Galametz19} are interpreted assuming a particular model for the temperature as a function of radius, which gives $\alpha_{\rm Pl}$. For the parameters of the sources considered in that paper, these model temperatures range from 20~K to 43~K at a radius of 500 AU.  This assumption about the temperature allows them to estimate $\beta$.
 \par
 We define $\beta_{1,3}$ to be the slope of the dust opacity between the frequencies corresponding to 1.3mm and 3.2mm.  This quantity, reported in \citet{Galametz19}, depends both on the dielectric function of the dust as well as on the size distribution.  For silicate dust, $\beta_{1, 3}$ is about 1.6 for a size distribution composed of grains much less than 1 mm in size, and becomes 0 for grains much larger than 3 mm in size.  
\par
$\beta_{1,3}$ is about 1.5 in the diffuse ISM \citep{Planck14}.  The values in \citet{Galametz19} at 500 AU range from 0.44 to 1.41, with a mean of 0.82, and are even lower at smaller distances from the star.  We can estimate how cold the emitting dust must be for these values of $\beta_{1, 3}$ to arise from an incorrect assumption about the temperature.  For a typical temperature of 30~K at 500~AU, $\alpha_{\rm Pl}^{1, 3} = 1.87$, where $\alpha_{\rm Pl}^{1, 3}$, analogous to $\beta_{1, 3}$, is the slope of the Planck function between the frequencies corresponding to 1.2 and 3.3 mm.  If in reality the temperature were lower, then $\alpha_{\rm Pl}^{1, 3}$ would be lower too, and the method of \citet{Galametz19} would underestimate $\beta_{1, 3}$.  However, to underestimate $\beta_{1, 3}$ by 0.7 due to temperature effects, the true value of $\alpha_{\rm Pl}^{1, 3}$ would have to be 1.17.  This would require a temperature of 6~K, an implausibly low number in protostellar environments.  The difference between 20 K and 43 K (the highest and lowest temperatures used in the modelling in \citet{Galametz19} at 500 AU) corresponds to a difference in $\alpha_{\rm Pl}$ of only 0.1.  For this reason, we do not consider low temperature to be a plausible explanation for the low spectral indices.
\par
  As can be seen from Equation \eqref{eq:fullAlpha}, a finite value of the optical depth will lower the observed $\alpha$.  However, it was shown in \citet{Galametz19} that the optical depths at 1.3 mm were only of order 0.1 at 500 AU.  This was deduced from $\tau_\nu = - \ln{(1-I_\nu/B_\nu)}$, by comparing the intensity $I_\nu$ of the observed emission with the intensity of a blackbody at the assumed temperature.  Again, if the true temperature were smaller by a factor of a few than the temperature they assumed, then their results would imply an optical depth of order unity, but this seems implausible.  Assuming $\beta = 1.5$, it would require an optical depth at 1.3 mm of 0.25 to reduce the inferred $\beta$ by 0.1, and of 2.2 to reduce the inferred $\beta$ by 0.7 (i.e., from 1.5 down to the 0.8 --- the mean of the observations at 500 AU).  

\subsubsection{Scattering}
In recent years it has been realised that as grains grow, the scattering cross section at millimeter wavelengths becomes important and cannot be neglected in high optical depth regions \citep[e.g. see the discussion in][]{Miotello22}. The effect of scattering may also modify the emission in high optical depth regions to appear as regions with moderate optical depth when interpreted in a classical fashion \citep[see e.g.][]{Zhu19}. It is important to realize that for scattering to be relevant, grains need to have grown to a significant fraction of the observing wavelength. This effectively implies that even considering the effects of scattering, when $T_d\ge 20$~K, a low measured value of $\beta_{1,3}$ associated with a moderate optical depth implies necessarily the presence of grain growth to sizes exceeding $100~\mu$m.

\subsubsection{Spinning dust emission}
    An additional possibility for the anomalous spectral index is contamination of the emission by something other than thermal emission.  In some of the studies motivating this work, it has been shown that the spectral indices between 1.3 and 3.2 mm can be contaminated by synchrotron radiation \citep{Galametz19}, and free-free emission \citep{Miotello14, Galametz19}.  In this section we consider also the possibility that electric dipole emission from spinning dust \citep{Draine98} could be a contaminant.
    \par
\begin{figure}  
\centering
\includegraphics[width=0.89\columnwidth]{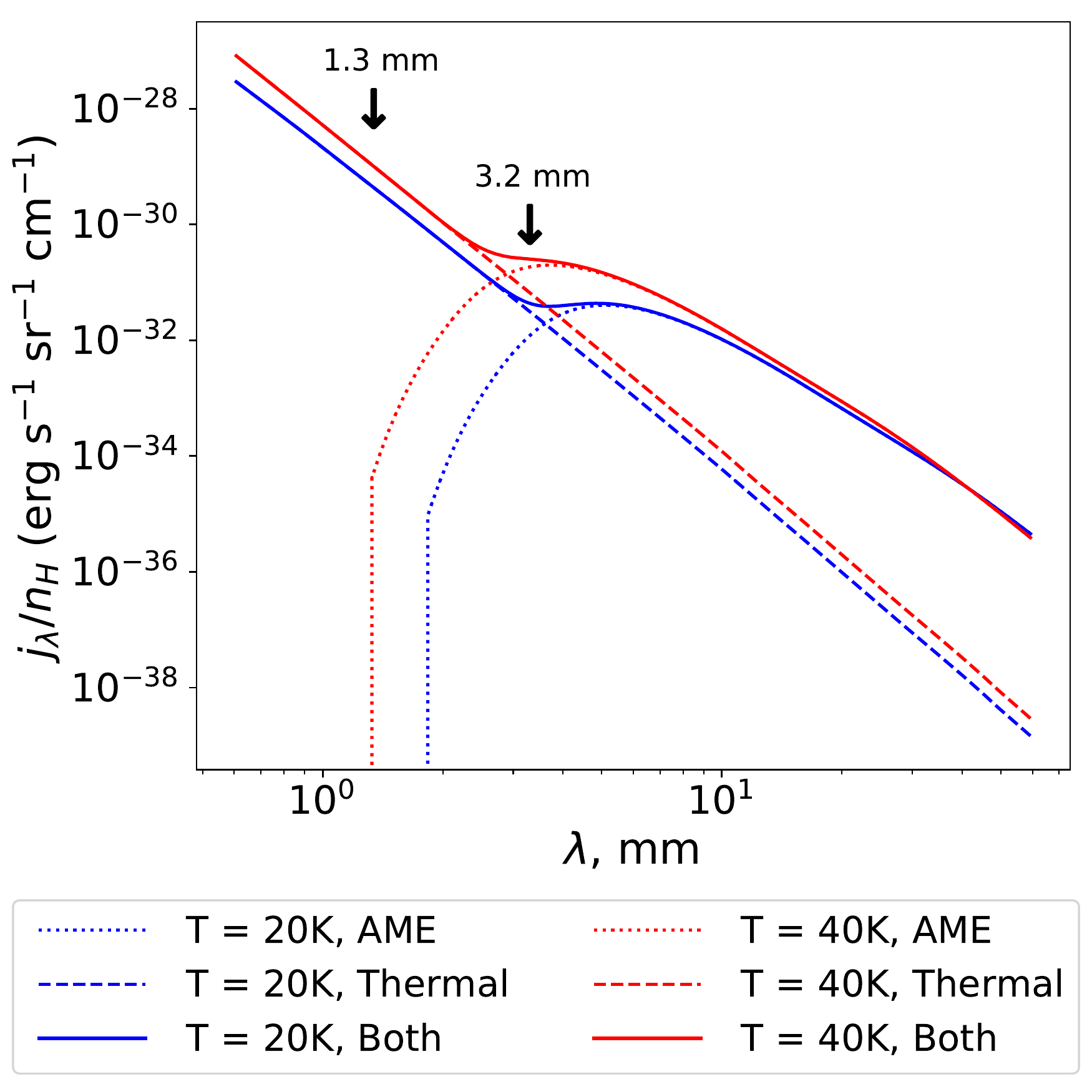}
\caption{Emissivity per hydrogen atom from thermal emission (dashed lines), spinning dust emission (dotted lines) and the combination (solid lines).  Blue curves correspond to an assumed gas temperature of 20~K and red curves to an assumed gas temperature of 40~K.  The arrows show the locations of the two wavelengths observed in \citet{Galametz19}.}
\label{fig:spinning}
\end{figure}

    To evaluate this, we used the SpDust code\footnote{\url{https://cosmo.nyu.edu/yacine/spdust/spdust.html}} \citep{AliHaimoud09, Silsbee11}.  We calculated the spinning dust emissivity at a hydrogen number density of $10^7$~cm$^{-3}$ and at temperatures of 20~K and 40~K.  We assumed the fiducial strength of the grain dipole moment from \citet{Silsbee11}.  The grain size distribution was assumed to be that given in line 16 of Table 1 of \citet{Weingartner01}. This corresponds to somewhat evolved dust with $R_V = 5.5$, however it still contains a negligible number of grains with sizes greater than 2 microns.  We stress that the spinning dust emission is dependent on the presence of nm-sized particles, whose abundance in such regions is very uncertain.  While simple modelling suggests that such small grains would have been removed from the size distribution very efficiently via coagulation \citep{Silsbee20}, PAHs have been detected in protoplanetary disks \citep{Seok17}.  Indeed, modelling in such disks \citep{Hoang18} shows that spinning dust emission can dominate over thermal emission at wavelengths $\gtrsim 3$~mm.  Assuming this grain size distribution, in Figure \ref{fig:spinning} we plotted the emissivity per hydrogen atom from thermal dust emission (dashed lines), spinning dust emission (dotted lines) and their combination (solid lines).  Red curves correspond to an assumed gas temperature of 40~K and blue to a gas temperature of 20~K.  The thermal dust emission was calculated using Mie theory, assuming the dust and gas temperatures to be equal.  The dielectric functions for silicate and carbonaceous grains are given in \citet{Draine03}.  We see that the spinning dust emission is completely irrelevant at 1.3 mm, but may substantially influence the observations at 3.2 mm.  The role of such emission could be clarified with more wavelength coverage, i.e. if observations were made with ALMA band 4 (as the longest wavelength where we expect no contribution from spinning dust), and in bands 1 and 2 (which would directly constrain the spinning dust contribution).
    \par
For $\nu_1$ and $\nu_2$ corresponding to emission at 1.3 and 3.2 mm, the spectral indices are $\alpha = 3.48$ and $\alpha = 3.58$ for $T = 20$~K and $T = 40$~K, respectively.  However, when we include also the spinning dust emission, these values drop to 3.22 and 2.28, respectively.  A value of $\alpha = 2.28$, interpreted as arising from thermal dust emission at 40~K, would imply  $\beta_{1, 3} = 0.4$, equal to the lowest of the measurements in \citet{Galametz19}.  That said, counter to the expectation if spinning dust emission were important, we note that there is essentially no correlation between the estimated temperature and the spectral index in the sources shown in Table 2 of \citet{Galametz19}.  In summary, we consider spinning dust an interesting possibility that cannot as of now be ruled out.

\section{conclusion}
Several recent observations of dust emission in the envelopes of young protostars have shown that the emission in the wavelength range between 1 - 3 mm has a spectral index much below what would be expected if the emission were dominated by thermal emission from small ($\ll$ 1 mm) dust grains.  This has been put forth as evidence for grain growth to millimeter sizes, but this interpretation is in tension with coagulation simulations that do not allow the coagulation of such grains in the relevant timescales.  Due to the large number of parameters controlling grain growth, it is difficult to make a definitive statement based on a tractable number of numerical simulations.  For that reason, we developed an analytic model of the grain growth and tested it against simulations.  
\par
The key result of this paper is given by Equation \eqref{eq:finalResult}, which shows the maximum ``optical equivalent" size $a_{\rm opt}$ a grain can grow to (for porous grains, this is the size of compact grain with similar optical properties).  This exact formula shows that $a_{\rm opt}$ only depends on the product of gas density $n_g$, threshold velocity $v_{\rm max}$ for grain fragmentation, and the lifetime $\tau_{\rm env}$ of the gaseous envelope.  For a threshold velocity of $10$~m/s, typical lifetimes of Class 0 protostars ($10^5$~years) and envelopes densities ($\leq 10^7$~cm$^{-3}$), this shows that grain growth cannot proceed to an equivalent size more than a few microns.  In order for mm-sized grains to form, this product would have to increase by 2 orders of magnitude, which we consider implausible.
\par
We consider other possible explanations for the low spectral index.  These include the effects of optical depth and temperature (which seem unlikely to contribute significantly) as well as possible (but poorly constrained) features in the dielectric function of the ices on the dust.  We also considered the possibility that the emission at long wavelengths is contaminated by electric dipole emission from the smallest dust grains.  This is plausible, but would require the presence of a large population of $\sim$nm sized grains.  It is also possible that large grains are present in the envelopes, but are not grown in situ. An attractive option could be the rapid growth in disks followed by extraction and redistribution in the envelope via powerful protostellar outflows  \citep{Wong16, Tsukamoto21}.
\\
\\
{\it Acknowledgments:}   We (KS, AI, MG, PC) gratefully acknowledge
the support of the Max Planck Society.  VA was supported by the grant of the Ministry of Science and Higher Education of the Russian Federation 075-15-2020-780 (N13.1902.21.0039). This work was partly supported by the Italian Ministero dell’Istruzione, Universit\`{a} e Ricerca through the grant Progetti Premiali 2012-iALMA (CUP C52I13000140001), by the Deutsche Forschungsgemeinschaft (DFG, German Research Foundation) - Ref no. 325594231 FOR 2634/2 TE 1024/2-1, by the DFG Cluster of Excellence Origins (www.origins-cluster.de). This project has received funding from the European Union’s Horizon 2020 research and innovation program under the Marie Sklodowska- Curie grant agreement No 823823 (DUSTBUSTERS) and from the European Research Council (ERC) via the ERC Synergy Grant ECOGAL (grant 855130).  The authors acknowledge Interstellar Institute's program ``The Grand Cascade" and the Paris-Saclay University's Institut Pascal for hosting discussions that nourished the development of the ideas behind this work.
\appendix
\section{Appendix A: Verification of assumptions}
\label{appendA}
\begin{figure}
\centering
\includegraphics[width=0.45\columnwidth]{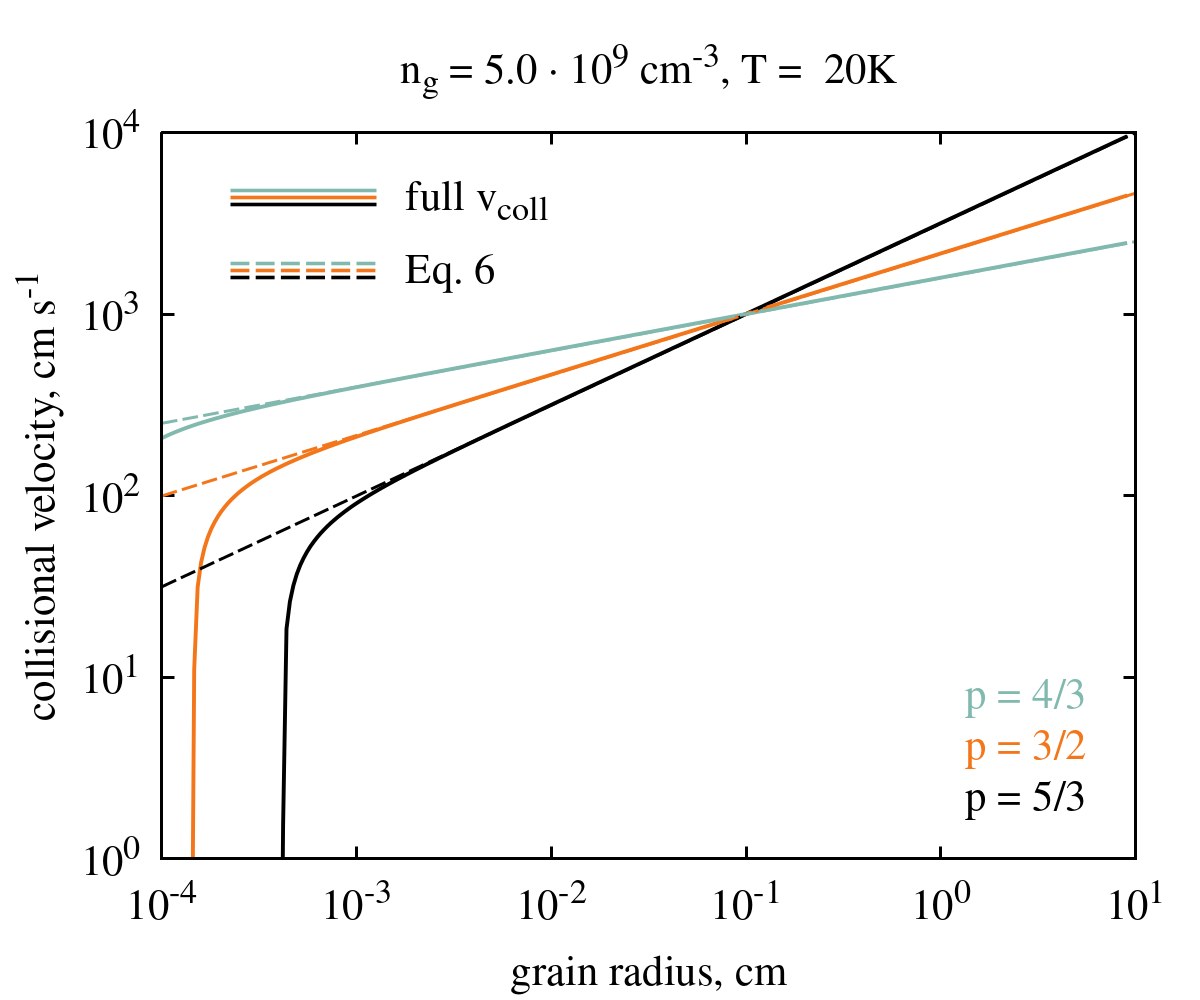}
\includegraphics[width=0.45\columnwidth]{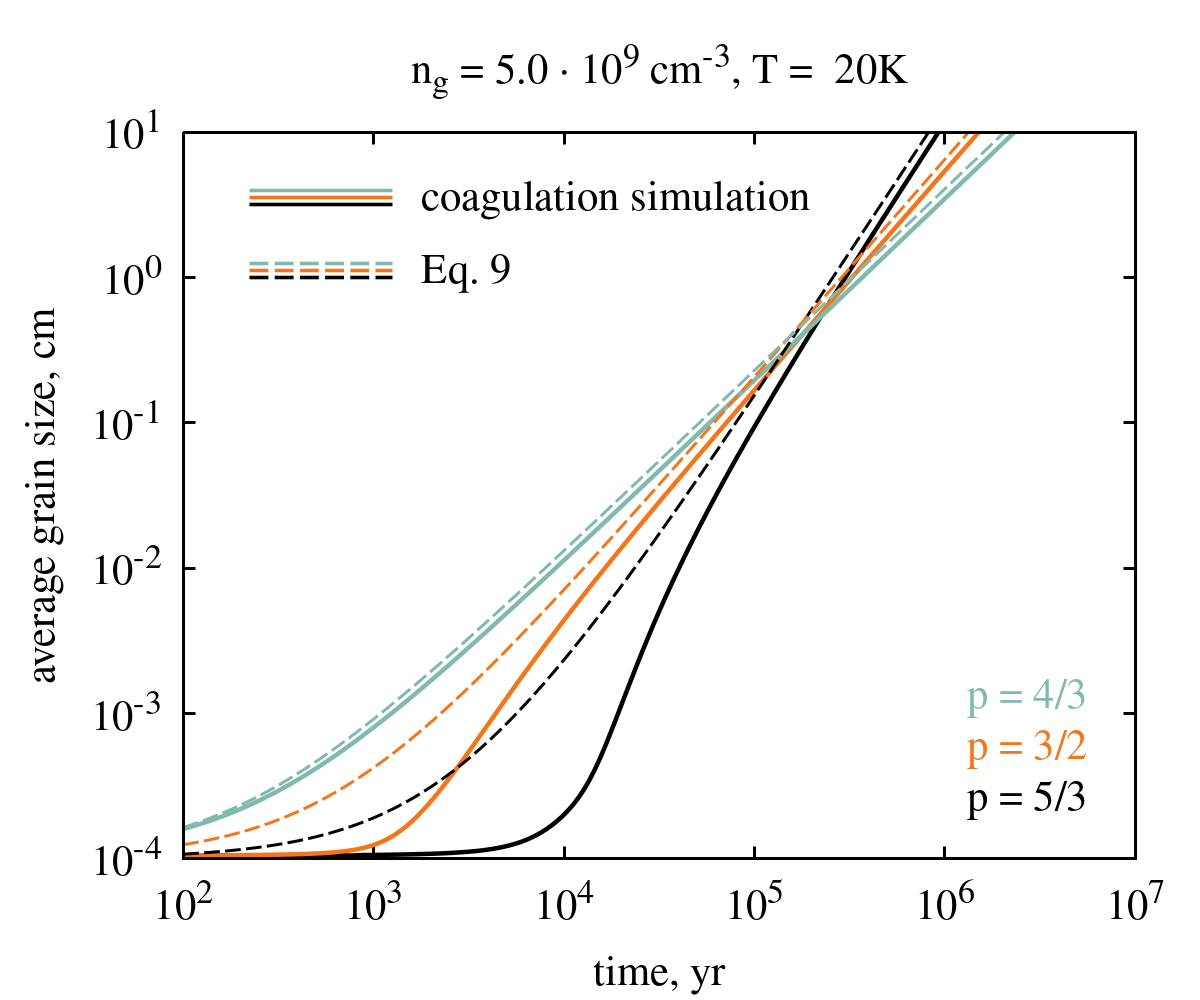}
\caption{Collisional velocities between grains of equal size (left) and average grain size evolution (right) for the numerical and analytical coagulation models (solid and dashed lines). The turbulent Mach number $M=v_{\rm g}/c_{\rm s}$ is set to $0.24, 0.51,$ and $1.17$ for MRI ($p=4/3$), Iroshnikov--Kraichnan ($p=3/2$), and Kolmogorov ($p=5/3$) turbulence, respectively, resulting in collision velocities of 10~m~s$^{-1}$ for 1~mm grains.}
\label{fig:bench_with_Smoluchowski}
\end{figure}

We verified with coagulation simulations that Equation \eqref{eq:growthEquation} provides a good approximation to the evolution of the dust size.  The left panel in Figure \ref{fig:bench_with_Smoluchowski} shows a comparison between a more accurate calculation of the collision velocities between grains of equal sizes made in \citet{Gong21}, and the analytic approximation made in Equation \eqref{eq:generalVelocity}.  The curves from \citet{Gong21} take into account both a dissipation scale for the turbulence, as well as Brownian motion.  These curves are essentially identical to those used in this paper for grains larger than 10 microns.
\par
The right panel of the plot shows a comparison between the average grain size found in the simulation, and that predicted by Equation \eqref{eq:growthEquation}.  The simulations were started with all grains having a size of $a_0 = 1~\mu$m, and assuming that grains remain compact with $D = 3$.
\par
The curves are different for two reasons.  First, the collision velocities used in the simulation are initially substantially smaller because of the assumed turbulence dissipation at small scales. This effect becomes unimportant once growth has proceeded well past the size at which the collision velocities in Equation \eqref{eq:generalVelocity} match those used in the simulation.  Second, our assumption of a monodisperse size distribution is not exactly correct, which could in principle affect the results at all sizes.  However, we see that in all cases the curves converge to much better than a factor of 2 at large sizes, where the collision velocities are the same.  
\par
We note that although the minimum eddy size is uncertain, this uncertainty can only serve to hinder the formation of large grains.  If the minimum eddy size were larger than used in the calculation of the left panel of Figure \ref{fig:bench_with_Smoluchowski}, then the cutoff (seen below $10^{-3}$~cm) would occur for larger sizes, and the coagulation would be further delayed.  This cutoff, provided it is not above the maximum grain size, will not affect the relationship between $v_g$ and $v_{\rm max}$.  Therefore, its presence can only act to slow down grain coagulation.

\bibliographystyle{apj}
\bibliography{Paper}
\end{document}